\documentclass[aip,jap,reprint,graphicx]{revtex4-1} 
\usepackage{graphicx,epsf,epsfig}
\usepackage{dcolumn}
\usepackage{bm}
\usepackage{amsmath, amsthm, amssymb}
\usepackage{color}
\usepackage{ulem}

\usepackage{epstopdf}
\usepackage{float}
\usepackage{longtable}
\usepackage{multirow}

\begin{document}

\title{Calculation of energy-barrier lowering by incoherent switching in STT-MRAM}
\author{Kamaram Munira}
\affiliation{Center for Materials for Information Technology, U. of Alabama, Tuscaloosa, AL 35401, USA}
\author{P. B. Visscher}
\affiliation{Center for Materials for Information Technology, U. of Alabama, Tuscaloosa, AL 35401, USA}
\affiliation{Department of Physics and Astronomy, Univ. of Alabama, Tuscaloosa, AL 35401, USA}

\begin{abstract}
To make a useful STT-MRAM (spin-transfer torque magnetoresistive random-access memory) device, it is necessary to be able to calculate switching rates, which determine the error rates of the device.
In a single-macrospin model, one can use a Fokker-Planck equation to obtain a low-current thermally activated rate $\propto \exp(-E_{eff}/k_B T)$.  Here the effective energy barrier $E_{eff}$ scales with the single-macrospin energy barrier $KV$, where $K$ is the effective anisotropy energy density and $V$ the volume.
A long-standing paradox in this field is that the actual energy barrier appears to be much smaller than this.
It has been suggested that incoherent motions may lower the barrier, but this has proved difficult to quantify.  In the present paper, we show that the coherent precession has a magnetostatic instability, which allows quantitative estimation of the energy barrier and may resolve the paradox.
\end{abstract}

\maketitle

\section{Introduction}
Most theoretical work on the switching of STT-MRAM\cite{MRAM} has used the single-macrospin model\cite{chap}\cite{ButlerEtal}.  This is adequate for very small elements in which the exchange interaction keeps the local magnetizations parallel, but when the volume $V$ is small, the stability parameter (energy barrier/$k_B T$, or $KV/k_B T$, where $K$ is the anisotropy energy density) is small.  For elements large enough to be stable, incoherent switching is possible.

One reason why it has been difficult to understand incoherent switching is that multi-macrospin switching simulations lead to a bewildering variety of motions -- precession can nucleate locally (perhaps in more than one place at the same time), precession or reversed domains can grow and shrink in an apparently random way, especially if the element is overdriven (\textit{i. e.}, the applied spin torque is much higher than the critical spin torque for onset of precession).  We have tried to simplify the problem by starting with the infinitesimal normal modes of oscillation about an initial uniform state, and continuing them to finite amplitude (Sec. \ref{section:norm}).

\section{Model}
We assume a cylindrical STT-MRAM element of thickness $t$ and radius $R$, with perpendicular anisotropy, stacked next to a pinned polarizing layer such that there is a spin torque proportional to the current in the LLG (Landau-Lifshitz-Gilbert) equation:
\begin{equation}\label{LLG}
\frac{d\mathbf{M}}{dt} = -\gamma \mathbf{M} \times \mathbf{H} - \frac{\gamma \alpha}{M_s}  \mathbf{M} \times  \mathbf{M} \times \mathbf{H}
- \frac{\gamma J}{M_s}  \mathbf{M} \times  \mathbf{M} \times \mathbf{\hat{m_p}}
\end{equation}
Here $ \mathbf{H}$ is the total field, including the exchange, anisotropy, and magnetostatic fields; $M_s$, $\gamma$, $\alpha$ are the saturation magnetization, gyromagnetic factor, and LL damping.  The coefficient $J$ of the spin torque is proportional to current, and has units of magnetic field (kA/m).
The anisotropy field is just $H_K M_z/M_s$, normal to the plane, where $H_K \equiv 2K/ \mu_0 M_s$.  The simulations in this paper were done with our public-domain micromagnetic finite-difference simulator\cite{alamag} -- the magnetizations are defined on a cubic lattice, the exchange field is a linear combination of neighboring magnetizations, and the magnetostatic field is computed using the Fast Multipole Method (FMM). We have omitted terms in $\alpha ^2$ since $\alpha$ is small.

\section{Normal modes }
\label{section:norm}
To study the statistics of switching, we must first characterize the initial state.  At zero temperature, this is the minimum energy state, which in an infinitely thin layer (or in a discretization with only one layer vertically) has magnetization vectors exactly along the normal (z) direction.  We will refer to it as the "flower" state, because in a system with nonzero thickness the magnetization near the tops of the edges bends outward (and inward at the bottoms).  At low temperatures, we will have only small fluctuations from this state, and the system can be characterized by a complete set of normal modes.  We have classified all of these normal modes and calculated many of them\cite{normodes}, but for the present purpose we need only the lowest-frequency modes, which have different symmetries that can be classified by an integer winding number $w$: the magnetization winds (about a vertical axis) $w$ times when we move around the element circumference once.  It turns out that these have the form\cite{normodes}
\begin{equation}\label{ansatz}
p(r,\theta) = ( \Re e^{iw\theta} r^w F(r), \Im e^{iw\theta}r^w F(r), 0)
\end{equation}
where F(r) is some smooth function.  Magnetostatic interactions make it impossible to calculate $F(r)$ analytically -- the best way we have found to compute the low-frequency normal modes is to start with a simple \textit{Ansatz} with the correct symmetry (Eq. \ref{ansatz} with F(r) = constant) and let the system evolve according to the LLG equation.  In the case of the lowest-frequency mode ($w=0$, a quasi-uniform state) the higher modes (which will initially be present with low amplitudes because the \textit{Ansatz} is not exact) will also have higher damping, and will gradually disappear, leaving the exact normal mode.  We keep the lowest mode from disappearing by re-normalizing it after each cycle, or by applying a spin torque (current) to counter the effects of damping.  The same can be done with the $w=1$ ("vortex") and $w=-1$ ("antivortex") modes, except that the known lower mode must be projected out to keep it from growing.

Of course, it is not possible to study switching using infinitesimal perturbations of the initial state.  We have been able to continue these normal modes to finite amplitudes, preserving the symmetry, by applying a current slightly higher than the critical value, so the amplitude drifts upward slowly.  The amplitude cannot be characterized by the precession angle $\theta$, as in a single-macrospin model, since the angle varies over the element, so we will characterize it by the total moment $m_z$ (which is $m_s \cos \theta$ in the single-macrospin model) instead.

\begin{figure}[!htb]
\begin{center}
\includegraphics[width=3 in]{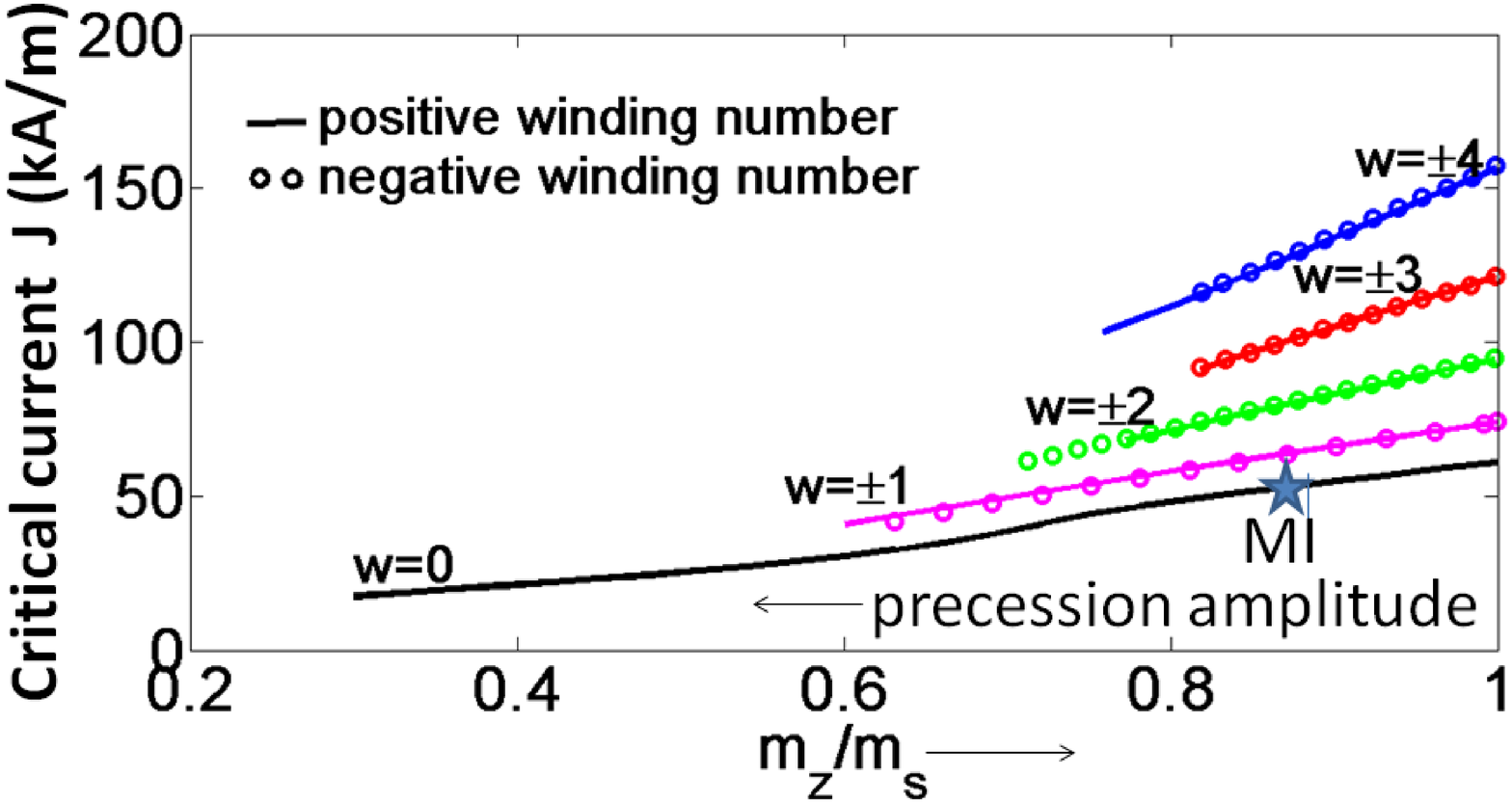}
\end{center}
\caption{\label{figure:Jc} "Critical current" J of normal modes, continued to finite amplitude by the numerical method described in the text, labeled by winding number (circles are positive $w$, line is negative $w$, but these seem to be nearly degenerate.)  Precession amplitude increases to the left; "MI" indicates magnetostatic instability.}
\end{figure}

The critical current to maintain precession (Fig. \ref{figure:Jc}) decreases as the precession amplitude increases (because the anisotropy field it must overcome decreases), so if the current is held constant the amplitude will increase uncontrollably.  Thus we must decrease the current slightly at each time to ensure that the precession grows slowly (quasi-statically).

The result of this process is a unique exactly periodic orbit for each amplitude (each $m_z$ in Fig. \ref{figure:Jc}).  The orbit can be continued past symmetry-breaking instabilities, by projecting the magnetization configuration onto the symmetry of the desired mode.  In particular, the coherent mode transforms as the $l=1$ representation of the rotation group\cite{normodes} while the instability discussed in the next section is a combination of $l=0$ and $l=2$, so by imposing $l=1$ symmetry, we can suppress the instability in Fig. \ref{figure:Jc}.

\section{Magnetostatic instability of quasi-uniform mode} \label{section:inst}
\begin{figure}[t]
\begin{center}
\includegraphics[width=3in, height=1.0 in]{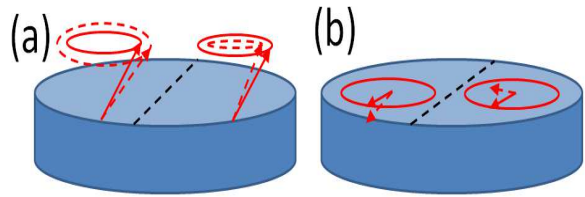}
\caption{\label{figure:MI} Cartoon of (a) quasi-uniform state (solid arrows and precession circle) and perturbed by largest-eigenvalue eigenvector (dashed arrows and precession circle), for small precession angle; (b) the same for $90^\circ $ precession angle, where instability is easier to understand.}
\end{center}
\vspace{-3mm}
\end{figure}

\begin{figure}[!htb]
\begin{center}
\includegraphics[width=3 in, height=2 in]{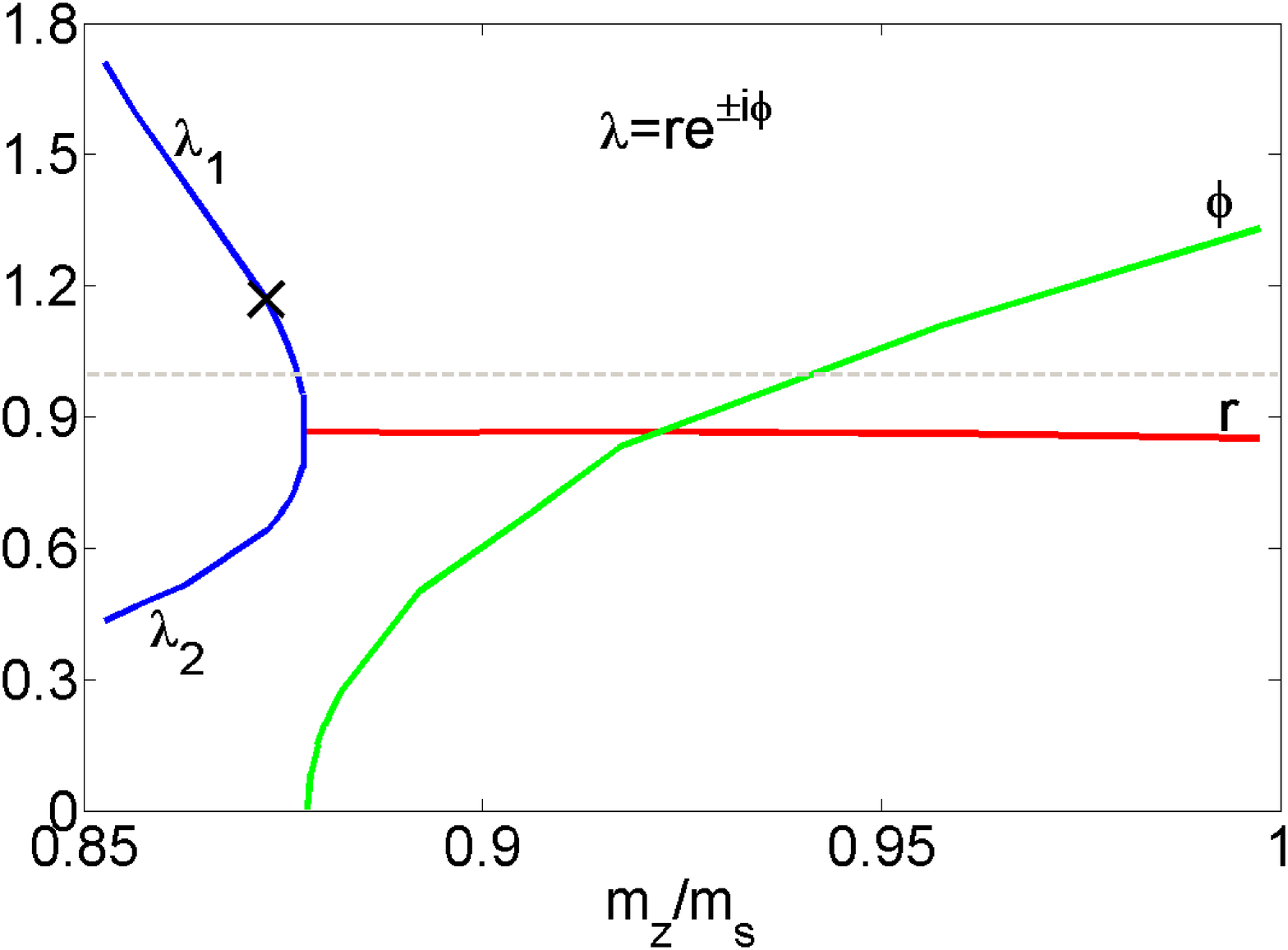} 
\end{center}
\vspace{-0.2 in}
\caption{\label{eig}Lyapunov eigenvalues vs. precession amplitude.  Since the matrix is real, they are either real ($\lambda_1$, $\lambda_2$) or complex conjugate pairs $\lambda_\pm = re^{\pm i \phi}$, in which case $r$ and $\phi$ are plotted.  Eigenvalue $\lambda_1=1.17$, whose eigenvector is shown in Fig. \ref{evol}(a), is marked with "x".  In these simulations $M_s = 500$ kA/m, $\alpha = 0.1$ for rapid convergence, $H_K = 1000$ kA/m, exchange $A = 10^{-11}$ j/m, $R = 30$ nm, $t = 4$ nm, cell size = 4 nm.}
\end{figure}

\begin{figure}[t]
\begin{center}
\includegraphics[width=3.2 in]{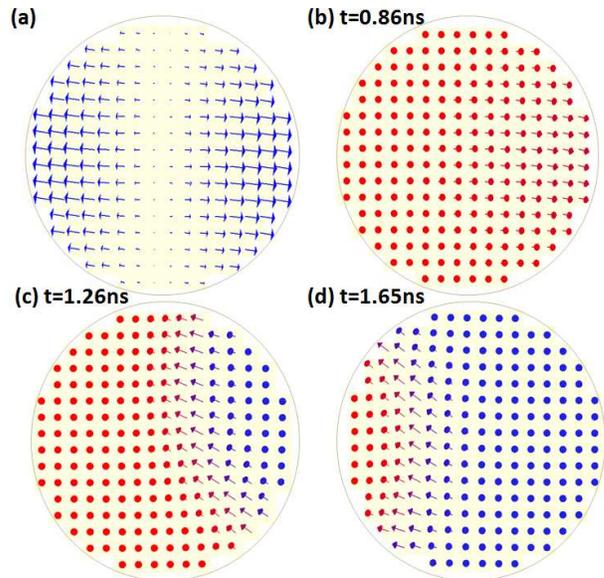}
\vspace{0.6 in}
\caption{\label{evol}
(a) Exact larger-eigenvalue eigenvector $p$ near the magnetostatic instability, at the point marked in Fig. \ref{eig} ($m_z/m_s= 0.8729$, $\lambda_1 = 1.17$). Addition of $\mathbf{p}$ to the rightward-tilting $\mathbf{M}_0$ increases the tilt at the right and decreases it at the left (b).  In (b-d), color of vectors encodes $M_z$ (positive, out of the paper, as on left, is red, into the paper is blue).  Subsequently the right edge reverses (c), forming a domain wall (with magnetizations precessing in the plane of the paper) that moves to the left (d).  Switching is complete when the wall reaches the left edge (not shown).  Only relative times are meaningful, because the initial amplitude is arbitrary.}
\end{center}
\end{figure}

The calculation leading to Fig. \ref{figure:Jc} is very stable for small amplitudes.  But for
larger amplitudes, there is a magnetostatic instability, shown schematically in Fig. \ref{figure:MI}(a).
One can see that there must be an instability before the magnetization becomes in-plane (Fig. \ref{figure:MI}(b), which shows the case $\theta = \pi/2$ with a perturbation in which the magnetization tilts upward at the right and downward at the left.)  This clearly lowers the anisotropy energy \textbf{and} the magnetostatic energy, analogously to stripe domains in an extended film\cite{fujiwara}, so clearly is unstable if exchange is weak.
The instability is also related to the Suhl instability of FMR precession in bulk systems\cite{suhl}.
It is not obvious at what angle this instability will occur, but we find numerically that it is unstable for $m_z$ below a critical value $m_{MI} \approx 0.875 m_s$ (Fig. \ref{eig}).
To study this instability, we first determine the exact "unperturbed" orbit $\mathbf{M}_0(\mathbf{r},t)$ for a specific amplitude ($m_z$), which is periodic with period $T$ (\textit{i. e., }$\mathbf{M}_0(\mathbf{r},t+T) = \mathbf{M}_0(\mathbf{r},t)$).
Then we add a perturbation $\mathbf{p}(\mathbf{r})$ and evolve $\mathbf{M}_0(\mathbf{r},0) + \mathbf{p}(\mathbf{r})$ for one cycle to some configuration $\mathbf{M}'(\mathbf{r})$, defining the evolved perturbation $\mathbf{p}(\mathbf{r}) \equiv \mathbf{M}'(\mathbf{r})- \mathbf{M}_0(\mathbf{r},T)$.  The map from $\mathbf{p}$ to $\mathbf{p'}$ is the Lyapunov map.  The system is stable if the eigenvalues of this map are all $< 1$.  Since this is a many-dimensional map, we can only determine its eigenvalues approximately.  One approach which works well is to assume that the eigenvectors with the largest eigenvalues are near the subspace spanned by the low-frequency modes (Eq. \ref{ansatz}) with $ w = 0, \pm 1$.  The eigenvectors with the same symmetry ($w=0$) as $M_0$ are easy to find -- one corresponds to shifting the phase of the orbit ($p = d\mathbf{M}_0/dt$) and the other to increasing the amplitude.  The $ w = \pm 1$ modes are linear in $x$ and $y$ -- it turns out that they mix, but if we use the basis $\mathbf{b}_{xmx}(x,y) = (x,0)$, $\mathbf{b}_{xmy}(x,y) = (0,x)$, $\mathbf{b}_{ymx}(x,y) = (y,0)$, $\mathbf{b}_{ymy}(x,y) = (0,y)$, in the first two ("$b_{xm}$") there is a vertical nodal line, to the left of which the magnetization tilts to the left (for $\mathbf{b}_{xmx}$) and to the right of which it tilts to the right.  The other two ("$b_{ym}$") have a horizontal nodal line, and don't mix with "$b_{xm}$".  Thus our 4x4 Lyapunov matrix $L$ is block-diagonal, and the two 2x2 blocks are identical by rotational symmetry.  Its elements are obtained by perturbing by a basis function $\mathbf{p} = \mathbf{b}_\beta (\beta = xmx$, etc): $L_{\alpha \beta} = \mathbf{b}_\alpha \cdot \mathbf{p'}$.  The eigenvalues are plotted in Fig. \ref{eig}.
The highest-eigenvalue eigenvectors can also be obtained by a more exact method (which does not assume it is in the 6D subspace described above.) If we start with an arbitrary perturbation $p$ and simply iterate the Lyapunov map (re-normalizing $p$ each time), components along eigenvectors with smaller eigenvalues will disappear, leaving us with the correct eigenvector, shown in Fig. \ref{evol}(a).

\section{Switching mechanism}
Armed with this understanding of the magnetostatic instability, we can describe the switching mechanism and predict the rate.  Clearly at high temperatures there will be many modes excited, so it is conceptually useful to consider the low-temperature limit in which the average energy $k_B T$ of the higher modes in a switching ensemble is much less than the energy in the quasi-uniform mode, which is biased in this ensemble to be near a switching energy $E_{sw}$.  Thus we will consider the limit in which the stability factor $E_{sw}/k_BT \rightarrow \infty$, although a realistic finite value probably behaves similarly.  Then in a switching trajectory, the quasi-uniform amplitude will perform a random walk, until $m_z \approx m_{MI}$, where the eigenvalue (Fig. \ref{eig}) passes 1.  At this point the system no longer evolves quasi-statically -- the eigenvalue $\lambda_1$ rises so quickly that the system will shoot out along the corresponding eigenvector, shown in Fig. \ref{evol}(a).  (Actually there are two degenerate eigenvectors differing by a $90^\circ$ rotation -- linear combinations produce nucleation at different points around the perimeter.) As long as this always leads to switching, the details may not matter.  We have looked at the evolution of this eigenvector -- except for an initial latency\cite{latency} (time lag), it is independent of the initial amplitude of the eigenvector -- thus the switching trajectory in this limit is essentially deterministic and unique (except for spatial rotations and time translations).  Fig. \ref{evol} shows several configurations along this trajectory\cite{mov} -- the left half of the system [where the perturbation (Fig. \ref{figure:MI}) narrows the precession cone] returns to the initial direction, but a reversed domain is formed in the other side, mostly at the edge.  This reversed domain expands by domain-wall motion, until the system is entirely switched.

\section{Activation energy and switching rate} \label{section:act}
Within the macrospin approximation, one can write a Fokker-Planck equation\cite{V&A,Zhang} for the evolution of the probability distribution.  In steady state the probability $\propto \exp(-E_{eff}/k_BT)$, where the effective energy\cite{V} satisfies
\begin{equation}\label{DE}
\frac{dE_{eff}(E)}{dE}  =  (1 - \frac{J}{J_c u})
\end{equation}
where we use the variable $u \equiv m_z/m_s = \cos \theta$ so we can generalize to a multi-macrospin system in which $\theta$ is not uniform.
Here $E = - KV u^2$, and $J_c = \alpha H_K$ is the critical current at which the initial state (with $\theta = 0, u=1$) is unstable against precession.
The solution to Eq. \ref{DE} is
\begin{equation}\label{Eeff}
E_{eff}/KV = 2 u J/J_c - u^2
\end{equation}

Determining the switching rate from the steady-state probability is not trivial\cite{rate}, but in the thermally activated regime it is proportional to the probability of being at the switching point $\theta_{sw}$, relative to the probability at the initial state: $\sim \exp(-E_b/k_BT)$ where the effective-energy barrier (activation energy) $E_b \equiv E_{eff}(u_{sw}) - E_{eff}({u=1})$.  In the macrospin model, $u_{sw} = u_{max}$, the value for which $E_{eff}$ is maximum, which is $J/J_c$, giving
\begin{equation}\label{Ebm}
E_b^{macrospin}/KV = (1 - J/J_c)^2.
\end{equation}

When we generalize to a multi-macrospin model, $E(u)$ won't change much -- the main change is that there is an instability.  We can construct a simple theory by keeping the single-macrospin $E_{eff}(u)$ but, if $u$ reaches $u_{MI}$ before it reaches $u_{max}$, i.e. $u_{MI} > u_{max}$, using $u_{sw} = u_{MI}$.  Then the energy barrier becomes
\begin{equation}\label{Eb}
E_b/KV = 1 - u^2_{MI} -2 (1-u_{MI}) J/J_c
\end{equation}

\begin{figure}[t]
\begin{center}
\includegraphics[width=3 in]{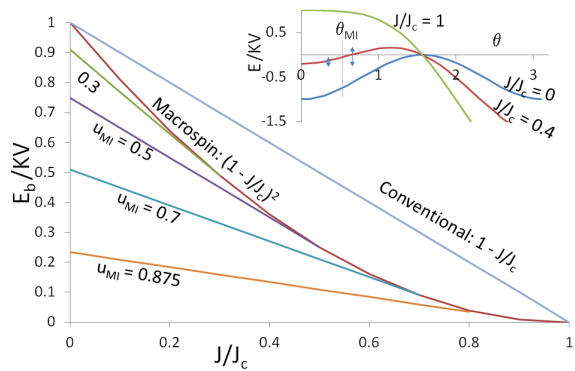} 
\caption{\label{barrier}The energy barrier as a function of spin-torque current, showing (from top) the conventional linear formula, the quadratic formula resulting from the single-macrospin theory, and the results of assuming various values for the instability threshold $u_{MI}$ -- each is a straight line, becoming tangent to the "macrospin" curve when the maximum in the energy landscape (inset) passes $\theta_{MI}$, an average angle at the instability defined by $\cos \theta_{MI}=u_{MI}$.}
\end{center}
\end{figure}

The energy landscape is shown in the inset to Fig. \ref{barrier}, which also shows the magnetostatic-instability angle $\theta_{MI} \approx 0.5$.
Clearly if we take $E_{sw}$ to be the instability-onset energy, the barrier (shown as double arrows in the inset) will be much smaller.
Several approximations to the barrier are shown in Fig. \ref{barrier}.  The parabola is the macrospin result; because experiments give a straighter line, often the exponent "2" is omitted and the straight line labeled "conventional" is used\cite{taniguchi}.  But also, the low-current energy barrier often appears\cite{lowbar} to be much less than $KV$, by a factor as small as $ \approx 1/6$; it has been suggested that $V$ should be replaced by a smaller "activation volume"\cite{actvol}.  However, it can be seen from Fig. \ref{barrier} that both of these problems (straightness and size) can be resolved by using the instability energy to determine the barrier.  In particular, the value $u_{MI} \approx 0.875$ we have found (the lowest line) gives a result close to the observed activation energies.  (Of course, this will depend on the particular parameters assumed, but can be calculated as in Sec. \ref{section:inst} above.)

\section{Conclusion}
In this paper we have presented a simple model for accounting for incoherence in STT-MRAM switching, which may resolve the problem that the observed activation energy is much lower than the single-macrospin prediction.  Our model assumes that incoherence can be neglected until the precession reaches a certain critical amplitude, at which a magnetostatic instability occurs and the system deterministically switches.  It is hoped that this simple theory can be improved by including the incoherent degrees of freedom explicitly.

\section{Acknowledgements}
This work was supported by Samsung Corporation.  We acknowledge useful conversations with Dmytro Apalkov and Alexey Khvalkhovskiy.

\end{document}